\documentstyle[prl,aps,epsfig,amssymb,floats]{revtex}
\begin{document}
\draft

\title{Curvature, hybridization, and STM images of carbon nanotubes}

\author{Alex Kleiner and Sebastian Eggert}
\address{Institute of Theoretical Physics \\ Chalmers
University of Technology and G\"{o}teborg University \\ 
S-412 96 G\"{o}teborg, Sweden \\ }
\date{Last change: \today}

\wideabs
{\maketitle
\begin{abstract}
The curvature effects in carbon nanotubes are studied analytically as a 
function of chirality.  The $\pi$-orbitals are found to be 
significantly rehybridized in all tubes, so that they are never
normal to the tubes' surface.
This results in a curvature induced gap in the electronic 
band-structure, which turns out to be larger than previous estimates.
The tilting of the $\pi$-orbitals should be observable by atomic resolution 
scanning tunneling microscopy measurements.
\end{abstract}

}

The electronic band-structure of carbon nanotubes has been a topic of 
intense investigation ever since their discovery in 1991\cite{iijima}.  
The basic electronic properties were quickly understood by
numerical studies of the graphite tight binding band structure together 
with a simple zone folding model\cite{hamada91,saito92}. 
In graphite the four outer electrons of carbon form  three
$sp^2$-hybridized $\sigma$-bonds and one $\pi$-orbital, which gives the 
conduction  band with six Fermi points and 
a linear dispersion around each of them\cite{wallace47}.
The electronic structure of the nanotube is then determined by the chiral 
wrapping vector along the direction $(n,m)$ 
since this determines whether the Fermi 
points satisfy the nanotube's circumferential boundary conditions.  
In that model, tubes with a chiral vector that satisfies ${\rm mod} [(n-m)/3] =0$
have their Fermi
point in the allowed $k$-space and thus are considered to be metallic,
while all other tubes are semiconducting\cite{hamada91,saito92}.
But even the ``metallic'' tubes may open a small gap if the bond symmetry 
is broken due to curvature\cite{buckbook,mintmire}, which has been analyzed 
in analytical studies in terms of a one-orbital tight binding 
approximation\cite{kane97,kleiner}.

One difficulty in predicting the effect of curvature on the 
electronic properties has been to determine the exact bond energies in
the curved graphite sheet to arrive at an analytical formula.   
So far it has explicitly been assumed that the $\pi$-orbitals are
orthogonal to the tubes surface\cite{kane97}, which is a common overly 
simplified picture that has also been used in a previous report by 
the authors\cite{kleiner}. 
We show in this Letter, however, that the $\pi$-orbitals are {\it never}
orthogonal to the surface, and instead are rehybridized due to 
the effect of the lower lying $\sigma$-bonds. 
Typically such a mixing effect is always
expected, but early studies have estimated that this band-mixing can be 
neglected\cite{saito92}.  However, for the case of metallic tubes 
we find that this mixing plays an important role, which is crucial in
determining an analytic formula for the curvature induced bandgap as a function
of chirality and curvature.  Moreover, we can predict the explicit angles
of the $\pi$-orbitals relative to the tubes surface, which should be observable
in atomic resolution pictures from Scanning 
Tunneling Microscopy (STM) experiments.

Our starting point is the well-defined geometrical structure of the 
carbon nanotubes by describing it in terms of an ``unrolled'' graphite sheet.  
The $\sigma$-bonds lie along three vectors which can be expressed 
in a coordinate system of the circumferential and 
translational axes $(\hat{c},\hat{t})$ in terms of the chiral indices $(n,m)$.

\begin{equation}\label{bond vectors}
\begin{array}{l}
\vec{R}_{1}=\frac{a}{2 c_{h}} 
[ (n+m)\hat{c} + \frac{1}{\sqrt{3}}(n-m)\hat{t} ]\\
\vec{R}_{2}=\frac{a}{2 c_{h}} 
[ -n\hat{c} + \frac{1}{\sqrt{3}}(n+2m)\hat{t} ]\\
\vec{R}_{3}=\frac{a}{2 c_{h}} 
[ -m\hat{c} - \frac{1}{\sqrt{3}}(2n+m)\hat{t} ],
\end{array}
\end{equation}
where $a \approx 2.49\rm \AA$ is the length of the honeycomb unit vector and
$c_{h}=\sqrt{n^{2}+nm+m^{2}}$ is the circumference in units of $a$.

\begin{figure}
\begin{center}
\mbox{\epsfig{file=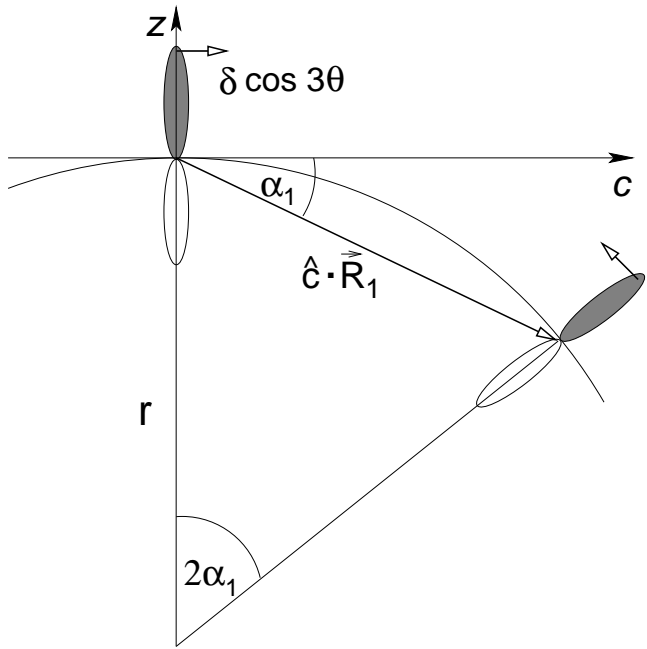,width=3.35in,angle=0}}
\end{center}
\caption{Schematic cross-section of a bond $\vec{R}_1$ in a 
nanotube with radius $r$.  The  $\pi$-orbitals are no longer normal 
to the tubes surface, but are titled by the 
hybridization angle $\delta =a/4\sqrt{3}r$ (small arrows). }
\label{sideview}
\end{figure}

In the regular $sp^2$ hybridization of the unrolled graphite sheet 
the four atomic wave functions can be written as 
\begin{eqnarray}\label{unhybrid}
|\sigma_i^0\rangle &=& \sqrt{\case{1}{3}}|s\rangle +
 \sqrt{\case{2}{3}}\left(\sin\beta_i |t\rangle +\cos\beta_i |c\rangle
 \right) ~~~~~~~~~i=1,2,3\nonumber\\
|\pi^0\rangle &=&|z\rangle
\end{eqnarray}
where 
$|s\rangle$ stands for the atomic s-orbital and $|t\rangle, |c\rangle,|z\rangle$ 
denote the 
p-orbitals along the translational, circumferential and normal 
directions in a nanotube, respectively.   
Here $\beta_i$ are the angles of the bonds relative to the 
circumferential direction ($\cos\beta_i \equiv  \hat c\cdot \hat R_i$).  
Each carbon atom has its own local 
coordinate system, where the $z$-direction is given by the normal direction 
to the graphite surface.
In a nanotube neighboring atoms have a relative angle $2 \alpha_i$ 
between their $z$-directions as shown in Fig.~\ref{sideview} with
\begin{equation}\label{sinalpha}
\sin \alpha_i = \frac{\hat c \cdot \vec{R}_i}{2r} = 
\frac{a}{2 \sqrt{3}r} \cos \beta_i,
\end{equation}
where $r = a c_h/2 \pi$ is the radius of the nanotube. 
We call this a {\it geometrical} tilting of the $\pi$-orbitals, 
which is known to induce a curvature gap\cite{kane97,kleiner} and
is predicted to cause a stretching  around the circumference 
of STM images\cite{lambin98}.

However, in addition we find an equally important contribution to the 
tilting from {\it hybridization}, which we will discuss next.
This hybridization comes from the fact that 
in a nanotube the three $\sigma$-bonds are not in the same plane, but
instead directed towards
the positions of the nearest neighboring carbon atoms, ie. 
they are tilted down by the angles $\alpha_i$
relative to the tangential $c$-direction as shown in Fig.~\ref{sideview}.
The hybridization of the $\sigma$-bonds is therefore changed 
from the uncurved expression 
for $|\sigma^0_i\rangle$ in Eq.~(\ref{unhybrid}) to
\begin{eqnarray}\label{hybrid}
& & |\sigma_i\rangle  = s_i |s\rangle \\ 
 & +& \sqrt{1- s_i^2}
\left(\,\sin\beta_i|t\rangle +
\cos\alpha_i \cos\beta_i|c\rangle-
\sin\alpha_i\cos\beta_i|z\rangle\,\right),\nonumber
\end{eqnarray}
where the mixing parameters $s_i$ (expanded 
around $s_i \approx \sqrt{1/3}$) can be determined by the three
orthonormality conditions between the $\sigma$ bonds 
$\langle\sigma_i|\sigma_j\rangle=\delta_{ij}$.  
The hybridized $\pi$-orbital can now be calculated in terms of the local 
basis of atomic orbitals by using the orthonormality conditions
\begin{equation}
\langle\pi|\sigma_i\rangle= 0.  \label{ortho}
\end{equation}

In what follows we will only work to lowest order in the curvature
parameter $a/r = 2\pi/c_h$ because energy band
repulsion will give higher order corrections 
for narrow nanotubes.  In particular, it 
has been shown  by ab initio LDA calculations that the
$\sigma$-bands are pushed up above the Fermi points for very small radii 
$r \alt 2.4\rm \AA$\cite{blase94}.  The straightforward 
hybridization analysis here is therefore only correct as a lowest order
approximation in $a/r\ll 1$, but gives a useful physical picture for most
observed nanotubes.

\begin{figure}
\begin{center}
\mbox{\epsfig{file=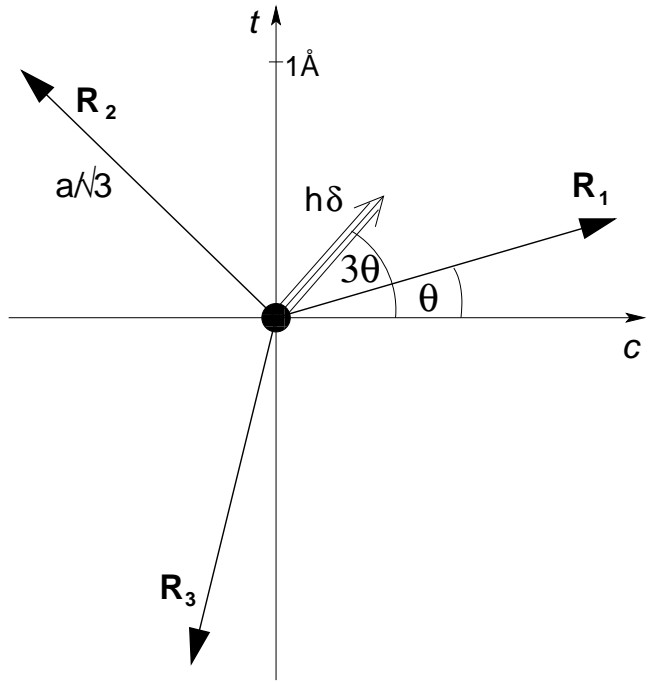,width=3.35in,angle=0}}
\end{center}
\caption{The three $\sigma$-bonds around a carbon atom in the 
rolled up graphite lattice. The unhybridized $\pi$-orbital 
is marked by the black dot in the center as seen from above.
The predicted hybridization in Eq.~(\ref{zhybrid}) 
of the $\pi$-orbital is indicated by the triple 
arrow with  $\delta = a/4\sqrt{3}r$.
The example shown here
corresponds to a (9,3) nanotube seen from a height of  
$h\approx 3 a$.
Note, that the angle $\theta$ in Eq.~(\ref{zhybrid}) can in fact be chosen 
to be any of the three angles $\beta_i$ of the bond vectors 
relative to the $c$-direction to get the correct hybridization.}
\label{bonds}
\end{figure}

\begin{figure}
\begin{center}
\mbox{\epsfig{file=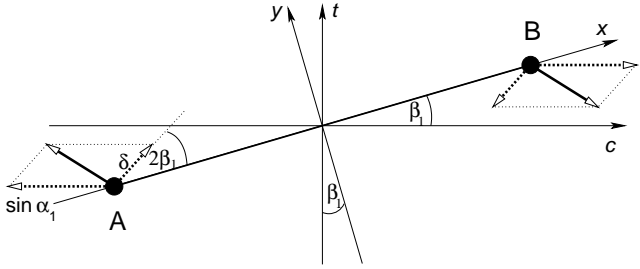,width=3.35in,angle=0}}
\end{center}
\caption{Two neighboring carbon atoms with 
with the predicted hybridization angle $\delta = a/4\sqrt{3}r$
and geometrical curvature effects $\sin\alpha_i = 2\delta \cos\beta_i$, 
seen from above. The example shown corresponds to the bond $\vec R_1$ in 
a (9,3) nanotube. } 
\label{topview}
\end{figure}

From Eqs.~(\ref{hybrid}) and (\ref{ortho}) we can now find the correct
expression for the $\pi$-orbitals to lowest order in $a/r$
\begin{equation}\label{zhybrid}
|\pi\rangle\simeq
|z\rangle+\frac{a}{4\sqrt{3}r}\left(\sqrt{2}|s\rangle\pm\sin
 3\theta|t\rangle\pm\cos 3\theta|c\rangle\right),
\end{equation}
where $\theta$ is the so-called chiral angle 
$\theta \equiv {\rm min}(|\beta_1|,
|\beta_2|,|\beta_3|), \ 0\leq \theta \leq \pi/6$ as shown in Fig.~\ref{bonds}. 
The different $\pm$ signs  in Eq.~(\ref{zhybrid}) refer to 
neighboring atoms $A$ and $B$ in the  bipartite graphite lattice,
since  their bonds and the corresponding angles $\beta_i$ point in opposite 
directions as indicated in Fig.~\ref{topview}. 

The physical interpretation of Eq.~(\ref{zhybrid}) is very intuitive.
The $\pi$-orbital is always inclined by the hybridization angle $\delta$ 
of size $\delta=a/4 \sqrt{3} r$
relative to the normal direction to the tube's surface.  However,
the direction of this inclination rotates with $3 \theta$
relative to the $c$-direction  as indicated in Fig.~\ref{bonds}.
It can be verified that the three angles between the $\pi$-orbital and 
each of the $\sigma$-bonds in Eq.~(\ref{hybrid}) are
equal as should be expected by symmetry.
Sometimes it is useful to express the hybridization in Eq.~(\ref{zhybrid}) 
in terms of the chiral indices $(n,m)$ using
\begin{eqnarray}\label{mn-formula}
\sin 3\theta &=&(n-m)(2m^2+5mn+2n^2)/2 c_h^3\\
\cos 3\theta &=&3 \sqrt{3} mn(m+n)/2 c_h^3 \nonumber
\end{eqnarray}

At this point we can proceed to calculate the hopping matrix elements $\gamma_i$
between neighboring $\pi$-orbitals which will determine the one-orbital
tight binding band structure. Following the Slater-Koster scheme\cite{slater54} 
for calculating the matrix elements $\gamma_i$ between 
tilted $\pi$-orbitals we need to use a common coordinate basis 
for the two neighboring atoms. 
Both the {\it hybridization} effect in Eq.~(\ref{zhybrid}) and 
the {\it geometrical} tilting with $\alpha_i$ in Eq.~(\ref{sinalpha})
can easily be expressed in the coordinate system 
where the bond vector $\vec{R}_i$
between two neighboring atoms $A$ and $B$ 
defines the x-direction as shown in Fig.~\ref{topview}  
\begin{eqnarray}\label{neighbors}
|\pi\rangle & \simeq &
|z\rangle
+ \sqrt{2}\delta |s\rangle
 \pm \left(\delta\cos 2 \beta_i - \sin\alpha_i \cos \beta_i\right) |x\rangle \\ & &
\pm\left(\delta \sin 2\beta_i + \sin\alpha_i \sin\beta_i\right)|y\rangle, \nonumber
\end{eqnarray}
where $\delta = a/4\sqrt{3}r$ and 
the angles $\beta_i$ are now defined in respect to the $A$ site.
Using the notation in Ref.~\onlinecite{mintmire}, we can use 
the overlap integrals between neighboring $s$- and 
$p$-orbitals $V_{ss\sigma}, V_{sp\sigma}, V_{pp\sigma}, V_{pp\pi}$ 
to calculate the hoping matrix elements $\gamma_i$ 
\begin{eqnarray}
& & \gamma_i = V_{pp\pi}\\ 
& -& \frac{a^2}{48 r^2}\left([3+8\sin^2 2\beta_i] V_{pp\pi}
- 2 V_{ss\sigma} - 2 \sqrt{2} V_{sp\sigma}  + V_{pp\sigma} 
\right).\nonumber
\end{eqnarray}
Here we have also used the second order term for the $|z\rangle$ orbital, 
which contributes to this expression.

The most interesting aspect of the electronic structure in metallic tubes 
is the size 
of the gap due to curvature as a function of the chiral wrapping vector, which
we can now calculate directly.
After having determined the rehybridized orbitals and hoping integrals,
we can typically ignore
the lower lying bands in further calculations and use a simple one-orbital
tight binding approximation. 
In this model the gap can be calculated from the positions
of the Fermi points $\vec{k}_F$ in the curved graphite sheet which in turn
are determined in terms of the new hopping matrix 
elements $\gamma_i$\cite{wallace47}
\begin{equation}\label{gap}
\sum_{i=1}^{3}\gamma_{i} e^{i\vec{k}_F\cdot\vec{R}_{i}}=0.
\end{equation}
Using the linear dispersion relation and the position of the quantization lines
it is then straightforward to derive the gap equation\cite{kleiner,yang}
assuming that the tube is metallic ${\rm mod}[(n-m)/3]=0$
\begin{equation}
E_g = \frac{2\sqrt{3}}{a}\left|\sum_{i=1}^3\, (V_{pp\pi}-\gamma_i)\, \vec{R}_i
\cdot \hat{t} \right| = \frac{a^2}{4 r^2} V_{pp\pi}\sin 3\theta.
\end{equation}
This surprisingly simple formula reconfirms again the notion that armchair
tubes $(n=m,\  \theta=0)$ do not have a gap from curvature, while 
zigzag-tubes $(m=0,\  \theta=\pi/6)$ and all other metallic tubes acquire
a gap of order $1/r^2$\cite{buckbook}.  For semiconducting tubes with an
intrinsic gap of order $1/r$ this geometrical and hybridization
correction to the gap can typically be neglected, but for metallic tubes 
it is essential to take the proper hybridization into account.
While the dependence of the gap on the
chiral angle agrees with previous analytic studies\cite{kane97,kleiner}, 
the expression in Eq.~(\ref{gap}) is by a factor of four larger than those 
estimates, which have not considered hybridization effects.

\begin{figure}
\begin{center}
\mbox{\epsfig{file=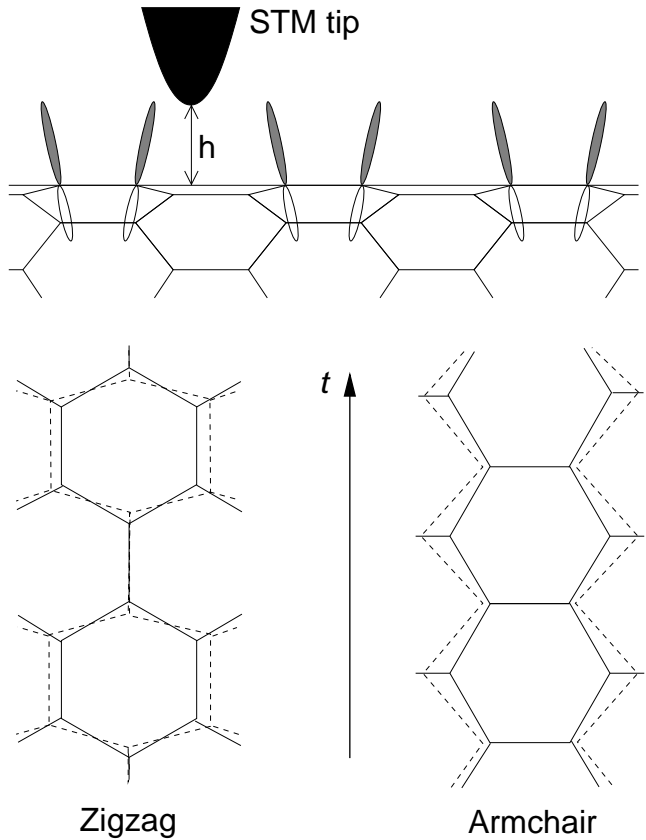,width=3.35in,angle=0}}
\end{center}
\caption{The predicted distortion of an STM image (dashed lines). 
For a zigzag tube we find
a pronounced squaring of the hexagons and a change of the lattice 
constants along the transverse directions.  For the armchair tube the
image is only affected along the circumferential direction. }
\label{stm}
\end{figure}

We now discuss how the rehybridization 
may be observable in STM experiments.  Already without rehybridization
the directions of the $\pi$-orbitals in a curved geometry 
can affect the STM images
as predicted in Ref.~\onlinecite{lambin98}.  In that study only the 
{\it geometrical} effects were taken into account, which resulted in 
an effective stretching of the STM image along the circumference by the
amount $x_c \to x_c (1+{h}/{r})$, where $h$ is the height of 
the STM tip relative to the nanotube surface.  However, the tilting $\delta$
caused by {\it hybridization} is of similar size 
and will results in an additional 
distortion of the STM image 
in both transverse and circumferential directions, but {\it alternating} for 
the inequivalent $A$ and $B$ atoms  
\begin{eqnarray} \label{distortion}
x_c &\to& x_c \left(1+\frac{h}{r}\right)\pm \frac{h a}{4\sqrt{3}r}
\cos 3\theta \nonumber \\
x_t &\to& x_t \pm \frac{h a}{4\sqrt{3}r} \sin 3\theta,
\end{eqnarray}
where $x_c$ and $x_t$ measure distances in the circumferential and
transverse directions, respectively.  Interestingly, the distortion from
the hybridization depends again to linear order on $h/r$ in all tubes, 
as opposed to the curvature gap which is a correction of second order
and only matters for metallic tubes.  If we consider this
distortion for the case of a zigzag tube with $\theta = \pi/6$, we see that
the hexagons of the graphite sheet will appear compressed in the
transverse direction, but stretched around the circumference 
as shown in Fig.~\ref{stm}.  For the armchair tubes $\theta =0$
on the other hand there is no deformation along the transverse direction,
but the hexagons are still deformed along the circumferential direction
as shown in Fig.~\ref{stm}.  
To observe the hybridization effects it would 
therefore be most advantageous to scan along the ridge of a zigzag tube
and average or Fourier transform the image 
over a distance of several hundred carbon sites, preferably for
more than one value of the height $h$.

In summary we have shown that it can be expected that the curvature of 
carbon nanotubes will result
in a significant rehybridization of the $\pi$-orbitals.  This rehybridization 
will affect the energy gap and will also manifest itself in a well-defined
distortion of atomic resolution STM images given by Eq.~(\ref{distortion}).  
To lowest order in the 
curvature parameter $a/r$ the hybridization 
angle was found to be $\delta = a/4\sqrt{3}r$, resulting in an 
energy gap which is significintly higher than previous analytical 
studies\cite{kane97,kleiner} as well as numerical estimates\cite{mintmire}. 
It is not clear to us what assumptions were used to model the curvature 
in the tight binding calculations of Ref.~\onlinecite{mintmire},
but the results coincide with analytic studies\cite{kane97,kleiner} 
that did not take any hybridization into account.
Higher order effects and energy band repulsion will
modify the size of the hybridization angle $\delta$ and therefore also the
energy gap. However, the direction
of the hybridization given by $3\theta$ relative to the 
circumference is correct as can be seen by symmetry arguments.
The exact size of the hybridization angle can most
reliably be found by analyzing STM images as outlined above, which 
in turn would lead to a more reliable estimate of the curvature gap.

\end{document}